\documentclass[aps,prd,groupaddress,tighten,nofootinbib,superscriptaddress,preprint]{revtex4}
\usepackage{amssymb}
\usepackage{amsmath,color}
\usepackage{epsfig}
\usepackage{graphicx,epsf,epsfig}
\usepackage{bbm}
\def\slc#1{\setbox0=\hbox{$#1$}           
    \dimen0=\wd0                                 
    \setbox1=\hbox{/} \dimen1=\wd1               
    \ifdim\dimen0>\dimen1                        
       \rlap{\hbox to \dimen0{\hfil/\hfil}}      
       #1                                        
    \else                                        
       \rlap{\hbox to \dimen1{\hfil$#1$\hfil}}   
       /                                         
    \fi}

\begin{document}
\renewcommand{\Re}{\text{Re}}
\renewcommand{\Im}{\text{Im}}
\title{Threshold effects on renormalization group running of neutrino parameters in the low-scale seesaw model}

\author{Johannes Bergstr{\"o}m}
\email{johbergs@kth.se}

\author{Tommy Ohlsson}
\email{tommy@theophys.kth.se}

\author{He Zhang}
\email{zhanghe@kth.se}

\affiliation{Department of Theoretical Physics, School of
Engineering Sciences, Royal Institute of Technology (KTH) --
AlbaNova University Center, Roslagstullsbacken 21, 106 91 Stockholm,
Sweden}

\begin{abstract}

We show that, in the low-scale type-I seesaw model, renormalization
group running of neutrino parameters may lead to significant
modifications of the leptonic mixing angles in view of so-called
seesaw threshold effects. Especially, we derive analytical formulas
for radiative corrections to neutrino parameters in crossing the
different seesaw thresholds, and show that there may exist
enhancement factors efficiently boosting the renormalization group
running of the leptonic mixing angles. We find that, as a result of
the seesaw threshold corrections to the leptonic mixing angles,
various flavor symmetric mixing patterns (e.g., bi-maximal and
tri-bimaximal mixing patterns) can be easily accommodated at
relatively low energy scales, which is well within the reach of
running and forthcoming experiments (e.g., the LHC).

\end{abstract}
\maketitle

\section{Introduction}\label{sec:introduction}

Experiments on neutrino oscillations have opened up a new window in
searching for new physics beyond the Standard Model (SM) during the
past decade. Since neutrinos are massless particles within the
renormalizable SM, one often extends the SM particle content in
order to accommodate massive neutrinos. Among various theories
giving rise to neutrino masses, the seesaw mechanism attracts a lot
of attention in virtue of its natural and simple description of tiny
neutrino masses. In the conventional type-I seesaw
model~\cite{Minkowski:1977sc,Yanagida:1979as,GellMann:1980vs,Mohapatra:1979ia},
three right-handed neutrinos are introduced with super-heavy
Majorana masses far away from the electroweak scale $\Lambda_{\rm
EW} = {\cal O} (100)~{\rm GeV}$. Small neutrino masses are strongly
suppressed by the ratio between the electroweak scale and the large
mass of the right-handed neutrinos, which on the other hand, leaves
the theory lacking in experimental testability. However, there are
alternatives that allow us to realize the seesaw mechanism at an
experimentally accessible level, e.g., the TeV scale. One popular
way to lower the seesaw scale is to introduce additional singlet
fermions, which have the same masses as the right-handed neutrinos
but different CP parity~\cite{Mohapatra:1986bd}. They can be
combined with the right-handed neutrinos to form four-component
Dirac fields, while the lepton number is broken by a small Majorana
mass insertion. In such a scenario, the masses of the light
neutrinos are strongly suppressed by the small Majorana mass
insertion instead of the seesaw scale, while the non-unitarity
effects in the lepton flavor mixing can also be boosted to an
observable level~\cite{Malinsky:2009gw}. Other possibilities like
structural
cancellation~\cite{Pilaftsis:1991ug,Kersten:2007vk,Zhang:2009ac,Ohlsson:2010ca}
or the minimal flavor seesaw
mechanism~\cite{Gavela:2009cd,Antusch:2009gn} may be employed to
construct low-scale seesaw models.

In principle, the current neutrino parameters are observed via
low-energy neutrino oscillation experiments. On the other hand, the
seesaw-induced neutrino mass operator is usually given at the seesaw
scale. Therefore, neutrino parameters are subject to radiative
corrections, i.e., they are modified by renormalization group (RG)
running effects. Typically, at energy scales lower than the seesaw
threshold (i.e., the mass scale of the corresponding seesaw
particle), the RG running behavior of neutrino masses and leptonic
mixing parameters should be described in an effective theory, which
is essentially the same for different seesaw models. However, at
energy scales higher than the seesaw threshold, full seesaw theories
have to be considered, and the interplay between the heavy and light
sectors could make the RG running effects particularly different
compared to those in the effective theory. However, in the spirit of
some grand unified theories (GUTs), a unified description of fermion
masses and flavor mixing depends on the lepton flavor structure at
the GUT scale, which inevitably requires the RG running between
seesaw particle thresholds and above. The full sets of
renormalization group equations (RGEs) in the
type-I~\cite{Minkowski:1977sc,Yanagida:1979as,GellMann:1980vs,Mohapatra:1979ia},
type-II~\cite{Magg:1980ut,Schechter:1980gr,Cheng:1980qt,Wetterich:1981bx,Lazarides:1980nt,Mohapatra:1980yp}
and type-III~\cite{Foot:1988aq} seesaw models have been derived, both in
the SM, and in the Minimal Supersymmetric Standard Model (MSSM)
\cite{Chankowski:1993tx,Babu:1993qv,Antusch:2001ck,Antusch:2001vn,Chao:2006ye,Schmidt:2007nq,Chakrabortty:2008zh}.
The general feature of the running parameters have also been
intensively studied in the literature, and it has been shown that
there could be sizable radiative corrections to the leptonic mixing
parameters at superhigh energy scales (see, e.g.,
Refs.~\cite{Bergstrom:2010qb,Ray:2010rz} and references therein). In
particular, certain flavor symmetric mixing patterns can be achieved
at the GUT scale indicating that there might exist some flavor
symmetries similar to the gauge symmetry (see, e.g.,
Ref.~\cite{Lin:2009sq} and references therein).

Since the RG evolution together with the threshold effects of the
heavy seesaw particles may result in visible corrections to the
leptonic mixing parameters, there is a need to look into the RG
running effects on the leptonic mixing parameters in the TeV seesaw
model. In this work, we will explore in detail the RG evolution of
neutrino masses and leptonic mixing parameters in the TeV type-I
seesaw model. In particular, we will show that the threshold effects
play a crucial role in the RG running of neutrino parameters, and
some phenomenologically interesting flavor symmetric mixing patterns
may be feasible even at an observable energy scale.

The remaining part of this work is organized as follows. In
Sec.~\ref{sec:RGE}, we first present general RGEs of the neutrino
parameters in the type-I seesaw model; in particular, the matching
conditions in crossing the seesaw thresholds. Then, the threshold
corrections to the neutrino parameters are discussed in detail in
Sec.~\ref{sec:threshold}. Based on the analytical results, we
illustrate a phenomenologically interesting numerical example in
Sec.~\ref{sec:numeric}, in which the bi-maximal leptonic mixing
pattern at TeV scale is shown to be compatible with experimental
data when the seesaw threshold corrections are properly taken into
account. Finally, a brief summary and our conclusions are given in
Sec.~\ref{sec:summary}. In addition, there are two appendices
including the detailed analytical treatment of the threshold
effects.

\section{RGEs in the type-I seesaw model}
\label{sec:RGE}

The simplest type-I seesaw model is constructed by extending the SM
particle content with three right-handed neutrinos $\nu_{\rm R} =
(\nu_{\rm R1},\nu_{\rm R2},\nu_{\rm R3})$ together with a Majorana
mass term of right-handed neutrinos. The mass part of the lepton
sector Lagrangian reads
\begin{eqnarray}\label{eq:Lmass}
-{\cal L} = \overline{\ell_{\rm L}} \phi Y_e e_{\rm R} +
\overline{\ell_{\rm L}} \tilde\phi Y_{\nu} \nu_{\rm R} + \frac{1}{2}
\overline{\nu^c_{\rm R}} M_{\rm R} \nu_{\rm R} + {\rm h.c.},
\end{eqnarray}
where $\phi$ is the SM Higgs boson, while $\ell_{\rm L}$ and $e_{\rm
R}$ denote lepton doublets and right-handed charged leptons,
respectively. Here $Y_e$ and $Y_\nu$ are the corresponding Yukawa
coupling matrices with $M_{\rm R}$ being the Majorana mass matrix of
the right-handed neutrinos. Without loss of generality, one can
always perform a basis transformation and work in the basis in which
$M_{\rm R}$ is diagonal, i.e., $M_{\rm R} = {\rm
diag}(M_1,M_2,M_3)$. At energy scales below the right-handed
neutrino masses, the heavy right-handed neutrinos should be
integrated out from the theory and the masses of the light neutrinos
are effectively given by a non-renormalizable dimension-five
operator
\begin{eqnarray}\label{eq:operator}
-{\cal L}^{d=5}_{\nu} = \frac{1}{2} (\overline{\ell_{\rm L}} \phi)
\cdot \kappa \cdot (\phi^T \ell^c_{\rm L}) + {\rm h.c.},
\end{eqnarray}
where the effective coupling matrix $\kappa$ can be obtained from
the type-I seesaw formula
\begin{eqnarray}\label{eq:kappa}
\kappa = Y_\nu M^{-1}_{\rm R} Y^T_\nu,
\end{eqnarray}
and the Majorana mass matrix of the left-handed neutrinos is
\begin{eqnarray}\label{eq:Mnu}
m_\nu \equiv \kappa v^2,
\end{eqnarray}
with $v \simeq 174~{\rm GeV}$ being the vacuum expectation value of
the Higgs field.

Because of the seesaw thresholds, the RG running of neutrino
parameters needs to be treated separately. At energies above the
seesaw thresholds, the full theory should be considered and the
relevant beta-functions are given
by~\cite{Chankowski:1993tx,Babu:1993qv,Antusch:2001ck,Antusch:2001vn}
\begin{eqnarray}\label{eq:RGE1}
16\pi^2 \mu \frac{{\rm d}Y_e}{{\rm d}\mu} & = &  \left( \alpha_e +
C^e_e H_e + C^\nu_e H_\nu\right)Y_e,
 \\
16\pi^2 \mu \frac{{\rm d}Y_\nu}{{\rm d}\mu} & = &  \left( \alpha_\nu
+ C^e_\nu H_e + C^\nu_\nu H_\nu \right)Y_\nu,
 \\
16\pi^2 \mu \frac{{\rm d}M_{\rm R}}{{\rm d}\mu} & = & C_{\rm R}
M_{\rm R} \left( Y^\dagger_\nu Y_\nu \right) + C_{\rm R}
\left(Y^\dagger_\nu Y_\nu\right)^T M_{\rm R}, \label{eq:RGE3}
\end{eqnarray}
where $H_f = Y_f Y^\dagger_f$ for $f=e,\nu,u,d$ and the coefficients
$(C^e_e,C^\nu_e,C^e_\nu,C^\nu_\nu,C_{\rm R}) =
(3/2,-3/2,-3/2,3/2,1)$ in the SM. The coefficient $\alpha_\nu$ is
flavor blind and reads
\begin{eqnarray}\label{eq:alpha}
\alpha_\nu  =  {\rm tr} \left(3 H_u + 3 H_d + H_e +  H_\nu
\right)-\left(\frac{9}{20} g^2_1 + \frac{9}{4} g^2_2 \right).
\end{eqnarray}
If we make use of $m_\nu$ at energy scales above the seesaw
threshold, we can derive from Eqs.~\eqref{eq:kappa} and
\eqref{eq:RGE1}-\eqref{eq:RGE3}
\begin{eqnarray}\label{eq:RGEm}
\frac{{\rm d}m_\nu}{{\rm d}t} \equiv \dot m_\nu  =  2 \alpha_\nu
m_\nu + \left( C^e_\nu H_e + C_m H_\nu \right) m_\nu + m_\nu \left(
C^e_\nu H_e + C_m H_\nu \right)^T,
\end{eqnarray}
with $C_m = 1/2$. Here, for simplicity, we have defined
$t=\ln\mu/(16\pi^2)$.

At energies below the seesaw thresholds, i.e., in the effective
theory, neutrino masses are attached to the dimension-five operator,
and the RGE of $\kappa$ reads
\begin{eqnarray}\label{eq:RGEkappaB}
\dot \kappa  =  \alpha_\kappa \kappa + \left( C^e_\nu H_e  \right)
\kappa + \kappa \left( C^e_\nu H_e \right)^T,
\end{eqnarray}
where
\begin{eqnarray}\label{eq:alphakappa}
\alpha_\kappa & = & 2{\rm tr} \left(3H_u+3H_d+H_e\right) + \lambda -
3g^2_2,
\end{eqnarray}
with $\lambda$ denoting the SM Higgs self-coupling constant.

\begin{figure}[t]
\begin{center}\vspace{0.cm}
\includegraphics[width=12cm]{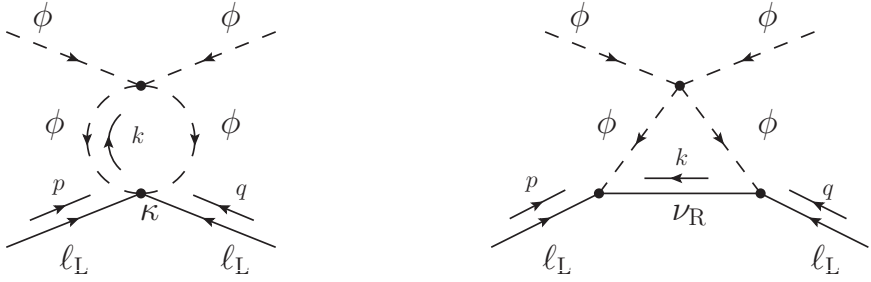}
\caption{\label{fig:fig1} Feynman diagrams of the one-loop $\lambda$
corrections to the neutrino mass operator in the effective theory
(left) and in the full theory (right), where $p,q$ and $k$ are the
corresponding momenta.} \vspace{0.cm}
\end{center}
\end{figure}

In crossing the seesaw thresholds, one should ensure that the full
and effective theories give identical predictions for physical
quantities at low energy scales, and therefore, the physical
parameters of both theories have to be related to each other. In the
case of the neutrino mass matrix, this means relations between the
effective coupling matrix $\kappa$ and the parameters $Y_\nu$ and
$M_{\rm R}$ of the full theory. This is technically called {\it
matching} between the full and effective theories. For the simplest
case, if the mass spectrum of the heavy singlets is degenerate,
namely $M_{1} = M_{2} =M_{3}=M_0 $, one can simply make use of the
tree-level matching condition at the scale $\mu=M_0$
\begin{eqnarray}\label{eq:match}
\kappa^{}_{} |_{M_0}=  Y_\nu M^{-1}_{\rm R}  Y^T_\nu |_{M_0}.
\end{eqnarray}
In the most general case with non-degenerate heavy singlets, i.e.,
$M_1 < M_2 < M_3$, the situation becomes more complicated and the
heavy singlets have to be sequentially decoupled from the
theory~\cite{Antusch:2002rr}. For instance, at energy scales between
the $n$-th and $(n-1)$-th thresholds, the heavy singlets are
partially integrated out, leaving only a $3 \times (n-1)$ sub-matrix
in $Y_\nu$, which is non-vanishing in the basis, where the heavy
singlet mass matrix is diagonal. The decoupling of the $n$-th heavy
singlet leads to the appearance of an effective dimension-five
operator similar to that in Eq.~\eqref{eq:operator}, and the
effective neutrino mass matrix below $M_n$ is described by two parts
\begin{eqnarray}\label{eq:m-eff}
m^{(n)}_\nu = v^2 \left[ \kappa^{(n)}+Y_{\nu}^{(n)}\left(M_{\rm
R}^{(n)}\right)^{-1}Y_{\nu}^{(n)T}\right],
\end{eqnarray}
where $(n)$ labels the quantities relevant for the effective theory
between the $n$-th and $(n-1)$-th thresholds. In the SM, the RGEs
for the two terms above have different coefficients for the gauge
coupling and Higgs self-coupling contributions, which can be traced
back to the decoupling of the right-handed neutrinos from the full
theory. For instance, we show in Fig.~\ref{fig:fig1} a comparison
between the $\lambda$ corrections in the effective theory (left
diagram) and in the full theory (right diagram).
The loop integral of the left diagram, i.e.,
\begin{eqnarray}
\int \frac{{\rm d}^4 k}{(2\pi)^4} \frac{1}{k^2 - m^2_\phi}
\frac{1}{(k-p-q)^2 - m^2_\phi} \nonumber
\end{eqnarray}
is UV divergent, whereas the loop integral of the right diagram,
i.e.,
\begin{eqnarray}
\int \frac{{\rm d}^4 k}{(2\pi)^4}\frac{\slc{k}+M_n}{k^2 - M^2_n}
\frac{1}{(k+p)^2 - m^2_\phi}\frac{1}{(q-k)^2 - m^2_\phi} \nonumber
\end{eqnarray}
is UV finite. Therefore, there is no $\lambda$ correction to the
neutrino mass operator in the full theory, while $\lambda$ enters
the beta-function of $\kappa$ in the effective theory. See also
Ref.~\cite{Antusch:2005gp} for a detailed discussion. We remark
that, if the relevant seesaw threshold is above the soft
SUSY-breaking scale, such a mismatch is absent in the MSSM due to
the supersymmetric structure of the MSSM Higgs and gauge sectors.
Therefore, this feature may result in significant RG running effects
only in the SM, in particular when the mass spectrum of the heavy
neutrinos is quite hierarchical.

\section{Threshold Effects in RGEs}
\label{sec:threshold}

We continue to determine the threshold corrections to the neutrino
parameters. According to Eq.~\eqref{eq:m-eff} the neutrino mass
matrix between the seesaw thresholds consists of two parts $\kappa$
and $ Y_\nu M^{-1}_{\rm R} Y^T_\nu $. The beta-functions of these
two parts have different coefficients in the terms proportional to
the gauge couplings and $\lambda$~\cite{Antusch:2005gp}. In the
framework of the SM, flavor non-trivial parts in the beta-functions
of $\kappa$ and $Y_\nu$ are dominated by $H_\nu$ and $H_e$, which
are essentially negligible in the RG running. For example, in a
basis where $H_e$ is diagonal, the largest entry of $H_e$ is
proportional to the square of the tau Yukawa coupling, i.e.,
$y^2_\tau$, which is at the order of $10^{-4}$ and too small to be
significant in the RG running. As for $H_\nu$, stringent constraints
from the unitarity of the leptonic mixing matrix indicate that their
contributions in the RG running are not comparable to the flavor
blind parts, i.e., the gauge couplings and $\lambda$. Hence, we
neglect the flavor non-trivial parts in the beta-functions in the
following analytical analysis.

For the RG running of the neutrino parameters from one seesaw
threshold $M_n$ to the nearby one $M_{n-1}$, the two parts of the
neutrino mass matrix are basically re-scaled to $a\kappa$ and $b
Y_\nu M^{-1}_{\rm R} Y^T_\nu $, with $a\neq b$. Explicitly, the mass
matrix of the light neutrinos at $\mu=M_{n-1}$ approximates to
\begin{eqnarray}\label{eq:mismatch}
m_\nu \big|_{M_{n-1}} & \simeq & b v^2 \left(\kappa +Y_\nu
M^{-1}_{\rm R} Y^T_\nu \right)\big|_{M_n} + (a-b)v^2 \kappa \big|_{M_n} \nonumber \\
& = & b \left[ m_\nu \big|_{M_{n}} + \left(ab^{-1}-1\right) v^2
\kappa\big|_{M_n} \right]  =  b \left( m_\nu \big|_{M_{n}} +
\varepsilon v^2  \kappa \big|_{M_n} \right),
\end{eqnarray}
where
\begin{eqnarray}\label{eq:a}
a &= & \exp \left(\int^{\frac{1}{16\pi^2}\ln
M_{n-1}}_{\frac{1}{16\pi^2}\ln M_n} \alpha_\kappa {\rm
d}t\right)\simeq \left(M_{n-1}\over M_{n}\right)^{
\frac{\alpha_\kappa}{16\pi^2}}, \\
b &= & \exp \left(\int^{\frac{1}{16\pi^2}\ln
M_{n-1}}_{\frac{1}{16\pi^2}\ln M_n} 2\alpha_\nu {\rm
d}t\right)\simeq \left(M_{n-1}\over
M_{n}\right)^{\frac{2\alpha_\nu}{16\pi^2}}, \label{eq:b}
\\
\varepsilon & = & \exp \left[\int^{\frac{1}{16\pi^2}\ln
M_{n-1}}_{\frac{1}{16\pi^2}\ln M_n} \left(\alpha_\kappa - 2
\alpha_\nu \right){\rm d}t\right] -1 \simeq
\left(\frac{M_{n-1}}{M_{n}}
\right)^{\frac{\alpha_\kappa-2\alpha_\nu}{16\pi^2}} -1.
\label{eq:eps}
\end{eqnarray}
In the limit $M_{n-1}=M_n$, one can easily observe that the second
term of Eq.~\eqref{eq:mismatch} disappears.

In the low-scale seesaw model, when the evolution of the neutrino
mass matrix crosses two thresholds (e.g., from $M_{n}$ to
$M_{n-1}$), the threshold corrections, i.e., the second term in
Eq.~\eqref{eq:mismatch}, provides corrections to the neutrino mass
matrix. In comparison to the RG running corrections, the threshold
effects are much more significant in changing the value of the
neutrino parameters. In order to see this point, we recall that when
the neutrino parameters are evaluated below the seesaw scale, i.e.,
in the effective theory, the only flavor non-trivial corrections in
the beta-functions are related to the charged lepton Yukawa coupling
$Y_e$. As we mentioned before, the largest entry in $Y_e$ is
constrained by the tau mass, and can only be at percentage level.
Therefore, no visible RG running effects can be gained in the
effective theory. At energies above the seesaw scale, $Y_\nu$
contributes to the RG running of the neutrino mass matrix as shown
in Eq.~\eqref{eq:RGEm}. In the ordinary seesaw models with GUT scale
right-handed neutrino masses, $Y_\nu$ can be sizable, e.g., at order
unity. However, in the low-scale seesaw model, $Y_\nu$ should be
relatively small in order to keep the stability of the masses of the
light neutrinos. Although there exist mechanisms that could
stabilize neutrino masses without the requirement of a tiny $Y_\nu$,
there is still very strong restrictions from the unitarity of the
leptonic mixing matrix. For example, if one of the right-handed
neutrinos is located around the electroweak scale $\Lambda_{\rm
EW}$, the unitarity of the leptonic mixing matrix sets a general
upper bound $H_\nu < 10^{-3}$ unless severe fine-tuning is involved.
Therefore, large RG running effects on the mixing parameters can
hardly be obtained above the seesaw scale. In the case of threshold
corrections, the situation becomes quite different, since in
Eq.~\eqref{eq:mismatch} the flavor non-trivial corrections are
proportional to the gauge couplings and $\lambda$, which do not
suffer from the unitarity constraints and can be of order unity. In
particular, if the light neutrino mass spectrum is highly
degenerate, sizable corrections to the neutrino parameters could be
naturally obtained.

In order to investigate the threshold effects on the neutrino
parameters, we present series expansions of the
leptonic mixing angles and the light neutrino masses. In
particular, we make use of the so-called Casas--Ibarra
parametrization~\cite{Casas:2001sr}, in which $Y_\nu$ is determined by
\begin{eqnarray}\label{eq:CI}
Y_\nu = U \sqrt{D_\kappa} O \sqrt{M_{\rm R}},
\end{eqnarray}
in a basis where $M_{\rm R}$ is diagonal. Here, $U$ is the leptonic mixing matrix $D_\kappa$ denotes
the eigenvalue matrix of $\kappa$, while $O =
R_{23}(\vartheta_{1})R_{13}(\vartheta_{2})R_{12}(\vartheta_{3})$
with $R_{ij}(\vartheta_k)$ being the elementary rotations in the
$23$, $13$, and $12$ planes, respectively. Different from the
leptonic mixing angles, $\vartheta_{i}$ are in general
complex.\footnote{See Ref.~\cite{Ellis:2005dr} for a discussion on
the role of the matrix $O$ in the RG evolution of the neutrino
parameters in the MSSM.} For simplicity, we will assume CP
conservation in this investigation (except from in the appendices),
i.e., $O$ is a real orthogonal matrix parametrized by three real
angles. A detailed derivation with CP-violating effects can be found
in Appendix~\ref{sec:app-A}. The neutrino mass matrix at the scale
$\mu = M_n$ can be diagonalized by using a unitary transformation,
viz.
\begin{eqnarray}\label{eq:U}
m_\nu \big|_{M_n}  = U ~{\rm diag} (m_1,m_2,m_3) ~U^{T},
\end{eqnarray}
with $m_i$ ($i=1,2,3$) being the light neutrino masses. Then, at the scale $\mu = M_{n-1}$, the
first-order correction to the leptonic mixing matrix is calculated
by
\begin{eqnarray}\label{eq:vector}
\tilde{U}_{\alpha j} = {U}_{\alpha j} + \varepsilon v^2 \sum_{k\neq
j} \frac {(U^T \kappa U)_{kj}}{m_j-m_k} U_{\alpha k} = {U}_{\alpha
j} +\varepsilon \sum_{k\neq j} \frac{\sqrt{m_j m_k} }{m_j-m_k}
O_{kn}O_{jn} U_{\alpha k}
\end{eqnarray}
and the masses of the light neutrinos are given by
\begin{eqnarray}\label{eq:mass}
\tilde{m}_i = b m_i + \varepsilon v^2
(U^T\kappa U)_{ii} =  \left( b + \varepsilon O_{in}^2 \right)m_i.
\end{eqnarray}
Here $n$ denotes the radiative corrections between the $n$-th and
$(n-1)$-th thresholds. See Appendix~\ref{sec:app-A} for the
detailed derivation of Eqs.~(\ref{eq:vector}) and (\ref{eq:mass}).

Inserting the parametrizations of $U$ and $O$ into the above
equations, one can easily obtain the threshold corrections to the
leptonic mixing parameters. For example, evolving the RGEs from $M_3$ to
$M_2$, the leptonic mixing angles are modified to
\begin{eqnarray}
\tilde{\theta}_{13} & \simeq &  \theta_{13}+\varepsilon
c_{{1}}c_{{2}} \sqrt {m_{{3}}} \left( {\frac
{s_{{1}}c_{{2}}s_{{12}}\sqrt {m_{{2}}}}{m_{{3}}-m_{{2}}}}+{\frac
{s_{{2}}c_{{12}}\sqrt { m_{{1}}}}{m_{{3}}-m_{{1}}}} \right), \label{eq:th13}\\
\tilde{\theta}_{12}  & \simeq & \theta_{12} +
\varepsilon s_1 s_2 c_2 \frac{\sqrt{m_1 m_2}}{m_2-m_1}, \label{eq:th12}\\
\tilde{\theta}_{23} &\simeq& \theta_{23} + \varepsilon
c_{{1}}c_{{2}} \sqrt {m_{{3}}} \left(- s_{{2}} s_{{12}}
\frac{\sqrt{m_{{1}}}}{m_{{3}}-m_{{1}}} + s_{{1}}c_{{2}} c_{{12}}
\frac{\sqrt{m_{{2}}}}{m_{{3}}-m_{{2}}}  \right) \label{eq:th23},
\end{eqnarray}
where $s_i =\sin\vartheta_i$ and $c_i=\cos\vartheta_i$, and the
terms proportional to $\varepsilon s_{13}$ have been ignored. The light
neutrino masses at the scale $\mu=M_2$ are approximately given by
\begin{eqnarray}
\tilde{m}_1 & = & \left( b + \varepsilon s_2^2 \right) m_1, \\
\tilde{m}_2 & =& \left( b + \varepsilon s_1^2 c_2^2 \right) m_2, \\
\tilde{m}_3 & =& \left( b + \varepsilon c_1^2 c_2^2 \right) m_3.
\end{eqnarray}
We refer the reader to Appendix~\ref{sec:app-B} for the complete
analytical derivation of the threshold corrections to the leptonic mixing
angles and the light neutrino masses.

Note that the mass differences in the denominator of
Eqs.~(\ref{eq:th13})-(\ref{eq:th23}) could strongly enhance the
threshold corrections. We give a rough estimate of the enhancement
effects. Unless in some fine-tuning limits, $\kappa v^2$ should
generally be the same order of magnitude as the neutrino masses,
i.e., $\kappa v^2 \sim m_\nu$. Taking into account the neutrino
mass-squared differences $\Delta m^2_{ij}=m^2_i- m^2_j$ measured by
neutrino oscillation experiments, one can calculate that, if $m_1
\sim 0.05~{\rm eV}$, $\kappa v^2 /(m_1-m_2) = {\cal O}(100)$ holds
no matter whether the mass hierarchy is normal or inverted. This is
a very distinctive feature of the RG running of the leptonic mixing
angles in the seesaw model, since such enhancement factors do not
exist in the quark sector. Due to the threshold effects, large
corrections to the leptonic mixing angles can be achieved even at a
very low energy scale, which provides us with a possibility of
testing the threshold effects experimentally. On the other hand, for
flavor symmetric theories constructed at low energy scales, e.g.,
the TeV scale, the seesaw threshold effects should be properly taken
into account although the RG running happens in a relatively short
distance.

Compared to the threshold corrections to the leptonic mixing angles,
there is however no enhancement factor in the RG evolution of the
neutrino masses. In the case of CP violation, the running
CP-violation phases are also affected by the threshold effects.
However, since there is no experimental information on leptonic CP
violation, we will not present further analysis of the running
CP-violating phases. In what follows, we will illustrate the
threshold effects on the leptonic mixing angles and concentrate on
the possibility of attaining flavor symmetric mixing patterns at a
low energy scale.

\section{numerical analysis of threshold effects in RGEs}
\label{sec:numeric}

In our numerical analysis, we adopt the values of the neutrino
mass-squared differences $\Delta m^2_{ij}$ and the leptonic mixing
angles $\theta_{ij}$ in the standard
parametrization~\cite{Amsler:2008zzb} from a global-fit of current
experimental data in Ref.~\cite{Schwetz:2008er} that we assume to be
given at the energy scale $\mu=M_Z$, where $M_Z$ denotes the mass of
the $Z$ boson. The values of the quark and charged-lepton masses as
well as the gauge couplings are taken from Ref.~\cite{Xing:2007fb}.
Furthermore, we take a representative value of the Higgs mass to be
$m_H = 140 ~{\rm GeV}$. For an illustrative purpose, the
right-handed neutrino masses are chosen as $(M_1,M_2,M_3) =
(200~{\rm GeV},500~{\rm GeV},1000~{\rm GeV})$. It is worth
mentioning that our choice of the right-handed neutrino masses
corresponds to $\varepsilon\sim -6 \times 10^{-3}$ between the
scales $M_3$ and $M_2$ and $\varepsilon\sim -8 \times 10^{-3}$
between the scales $M_2$ and $M_1$, cf.~Eq.~(\ref{eq:eps}).

\begin{figure}[t]
\begin{center}\vspace{-0.4cm}
\includegraphics[width=10cm]{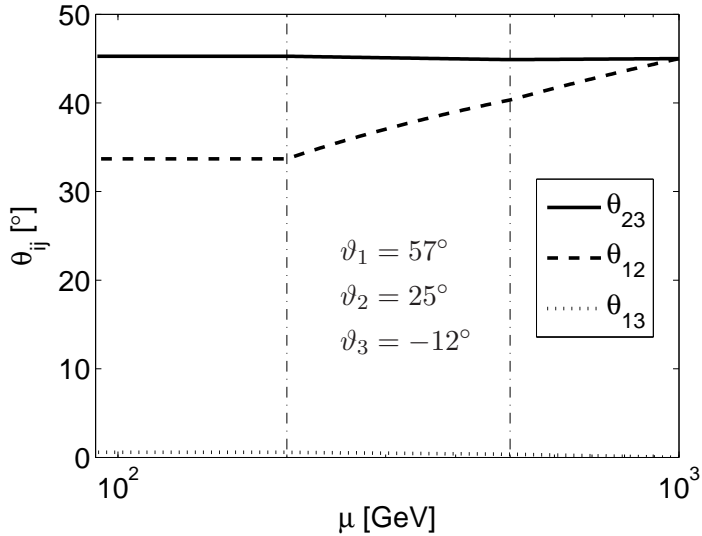}
\put(-155,105){$\vartheta_{1}=57^\circ$}
\put(-155,88){$\vartheta_{2}=25^\circ$}
\put(-155,71){$\vartheta_{3}=-12^\circ$} \caption{\label{fig:fig2}
The RG evolution of the leptonic mixing angles $\theta_{23}$,
$\theta_{12}$, and $\theta_{13}$ with solid, dashed, and dotted
curves, respectively. The threshold of the right-handed neutrinos
are marked by dashed-dotted vertical lines. At the scale $M_3 = 1
~{\rm TeV}$, the bi-maximal mixing pattern, i.e.,
$\theta_{12}=\theta_{23}=\pi/4$ together with $\theta_{13}=0$, is
assumed, while the corresponding rotation angles in $O$ are
given in the plot. In addition, the light neutrino mass spectrum is
taken to be normal hierarchy with $m_1(M_3)=0.05~{\rm eV}$. Because
of the lacking of experimental evidence of leptonic CP violation, we
assume all CP-violating phases to be zero in our calculation.}
\vspace{0cm}
\end{center}
\end{figure}

In Fig.~\ref{fig:fig2}, we show one example of the threshold effects
on the leptonic mixing angles.
The rotation angles in $O$ at the scale $\mu = M_3$ have been
labeled in the plot. In addition, we use the normal mass hierarchy,
i.e., $m_1<m_2<m_3$ together with $m_1(M_3)=0.05~{\rm eV}$, which
translates into $m_1 \simeq 0.047 ~{\rm eV} $ at the $M_Z$ scale.
One observes that, with the threshold effects included, the exact
bi-maximal mixing pattern (i.e., $\theta_{12}=\theta_{23} =
45^\circ$ and $\theta_{13}=0$) at a very low energy scale could be
compatible with experimental measurements. In particular,
$\theta_{12}$ decreases by about ten degrees from $M_3$ to $M_1$.
This can be compared to a similar study of obtaining the bi-maximal
mixing pattern at the GUT scale~\cite{Antusch:2002hy}. Such a
significant effect can be understood from our analytical results.
For instance, in the RG running between $M_3$ and $M_2$, we have
$s_1 s_2 c_2 = \sin(57^\circ)\sin(25^\circ)\cos(25^\circ) \simeq 0.3
$, $\varepsilon \simeq -6 \cdot 10^{-3}$ as well as $\sqrt{m_1
m_2}/(m_2-m_1) \simeq 43$ (at $\mu = M_3$), which leads to
a correction $\theta_{12}' \simeq -4.7^\circ$ according to Eq.~\eqref{eq:th12}
[or Eq.~(\ref{eq:theta'12})]. Continuing the RG running between
$M_2$ and $M_1$ [see Eq.~(\ref{eq:theta'12_2})], we obtain a total
correction $\theta'_{12} \simeq -9.2^\circ$ between $M_3$ and $M_1$, which
agrees approximately with the numerical value that is given by
$\theta'_{12} \simeq -12.5^\circ$. In general, varying the value of
the smallest light neutrino mass as well as changing the values of
the Casas--Ibarra angles, we observe that the analytical results,
which are only accurate for small corrections, are in very good
agreement with the numerics.

Furthermore, a non-trivial $\theta_{13}$ can be expected at the
$M_Z$ scale because of the threshold corrections, although we have
assumed $\theta_{13}=0$ at high energy scales. Apart from the
bi-maximal mixing pattern, other attractive flavor symmetric
structures, e.g., the tri-bimaximal mixing pattern (i.e.,
$\theta_{12}\simeq 35.3^\circ$, $\theta_{23} = 45^\circ$ and
$\theta_{13}=0$)~\cite{Harrison:2002er,Harrison:2002kp,Xing:2002sw}
can also be easily obtained through a proper choice of the rotation
angles $\vartheta_i$ (see also
Refs.~\cite{Mei:2004rn,Plentinger:2005kx,Luo:2005fc,Hirsch:2006je,Dighe:2006sr,Dighe:2007ksa,Boudjemaa:2008jf,Varzielas:2008jm,Goswami:2009yy}
for discussions on possible RG corrections to the tri-bimaximal
mixing pattern in other frameworks).

Note that, although all the three mixing angles do not stay at their
initial input values, the change of $\theta_{12}$ is generally
larger than those of the other two mixing angles. In view of
Eqs.~(\ref{eq:th13})-(\ref{eq:th23}), $\theta_{13}$ and
$\theta_{23}$ could only be amplified by the inverse of $\Delta
m^2_{32} = m_3^2 - m_2^2$ or $\Delta m^2_{31} = m_3^2 - m_1^2$, but
not the inverse of $\Delta m^2_{21} = m_2^2 - m_1^2$, since these
angles are located in the third column of the standard
parametrization. As we have shown in the previous section, the
boosting effect of $\Delta m^2_{21}$ is much more significant than
those of $\Delta m^2_{32}$ and $\Delta m^2_{31}$. Therefore,
$\theta_{12}$ receives a more significant RG correction. In
principle, if the mass spectrum of the light neutrinos is highly
degenerate, visible changes of all the leptonic mixing angles can be
expected.

It should be mentioned that, in the supersymmetric framework, the RG
running of leptonic mixing angles may also be enhanced by a large
value of $\tan\beta$. However, it has been shown that, in the MSSM,
$\theta_{12}$ generally tends to increase when running down from
high energy scales~\cite{Miura:2002nz,Antusch:2003kp}, which means
that many flavor symmetric mixing patterns are unfavorable.

\begin{figure}[t]
\begin{center}\vspace{-0.4cm}
\includegraphics[width=10cm]{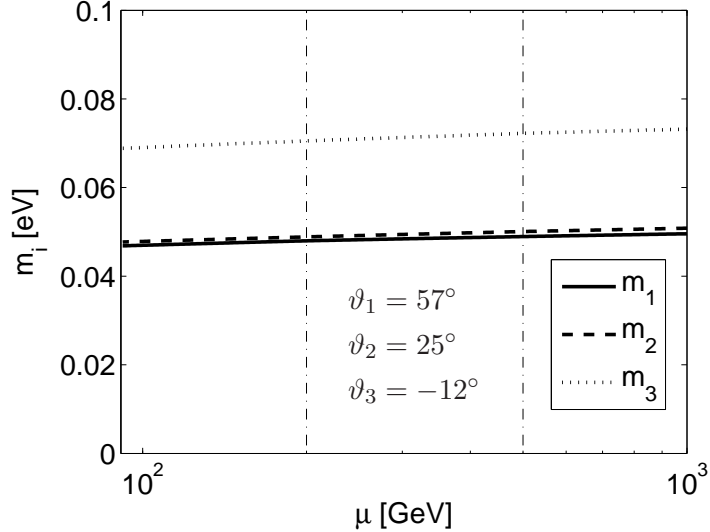}
\put(-155,85){$\vartheta_{1}=57^\circ$}
\put(-155,68){$\vartheta_{2}=25^\circ$}
\put(-155,51){$\vartheta_{3}=-12^\circ$} \caption{\label{fig:fig3}
The RG evolution of the masses of the light neutrinos $m_1$, $m_2$,
and $m_3$ with solid, dashed, and dotted curves, respectively. The
corresponding input parameters are the same as those in
Fig.~\ref{fig:fig2}.} \vspace{0cm}
\end{center}
\end{figure}

For completeness, we also show the RG running of the masses of the
light neutrinos in Fig.~\ref{fig:fig3}.
One reads directly from the figure that there is no significant
threshold corrections to the neutrino masses, which is consistent
with our analytical results. Note that we do not include
CP-violating effects in our analysis. Now that they may affect the
RG running behavior of mixing angles, a thorough study of the RG
running CP-violating phases is worthy, but out of the scope of this
work, and hence will be elaborated elsewhere.

\begin{figure}[t]
\begin{center}\vspace{0.5cm}
\includegraphics[width=5cm,bb=150 100 880 750]{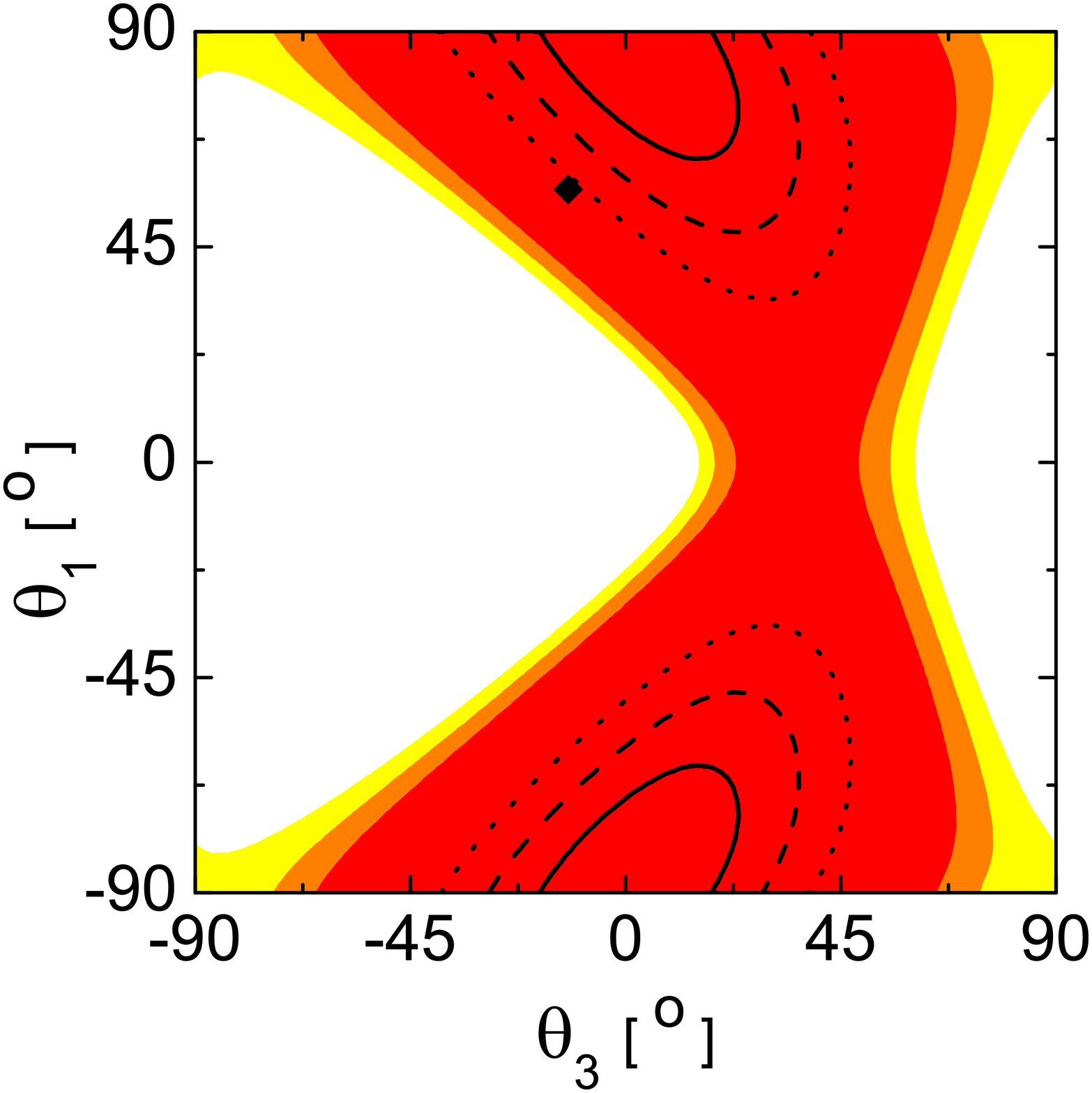}
\includegraphics[width=5cm,bb=80 100 810 750]{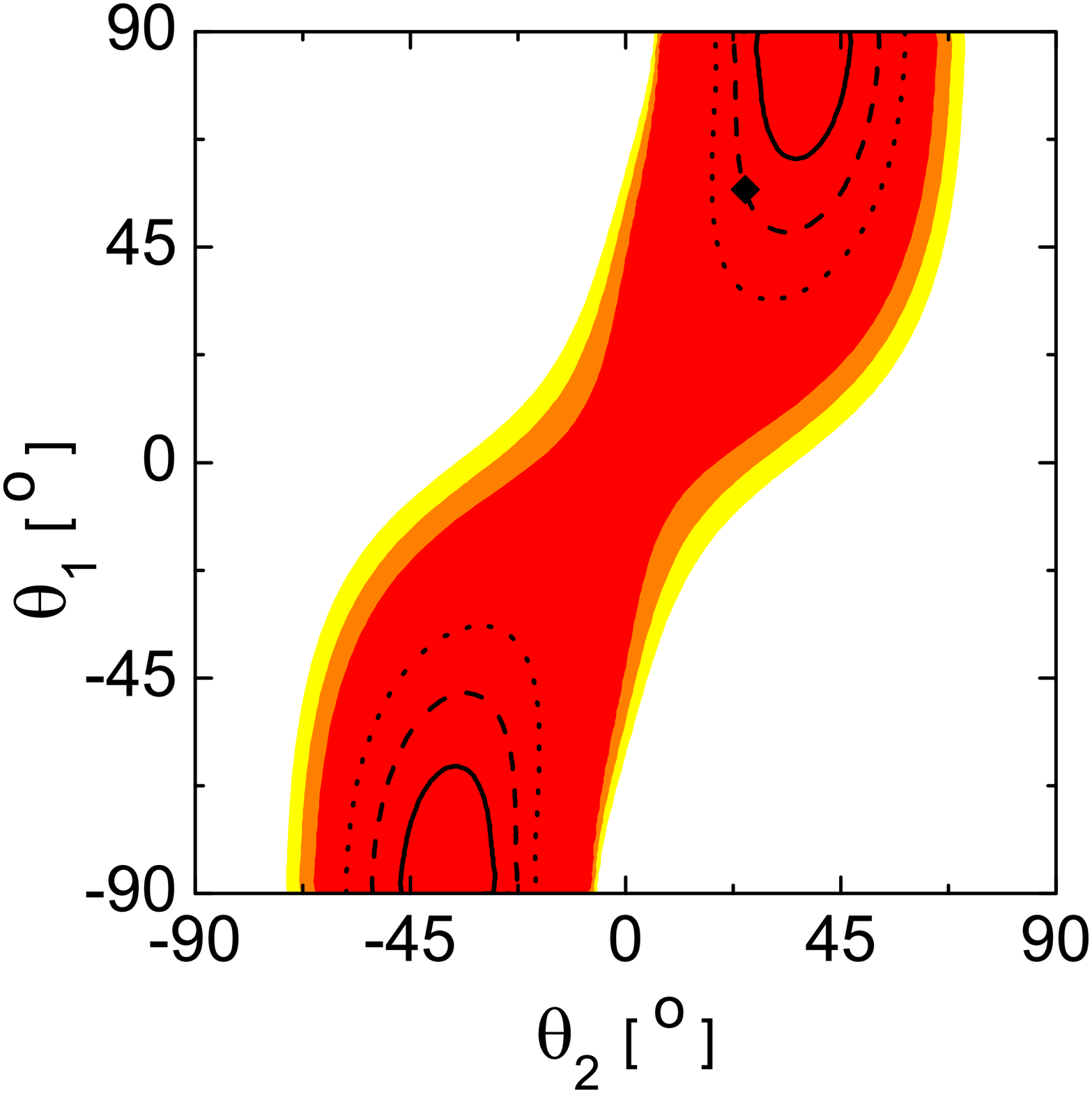}
\includegraphics[width=5cm,bb=10 100 740 750]{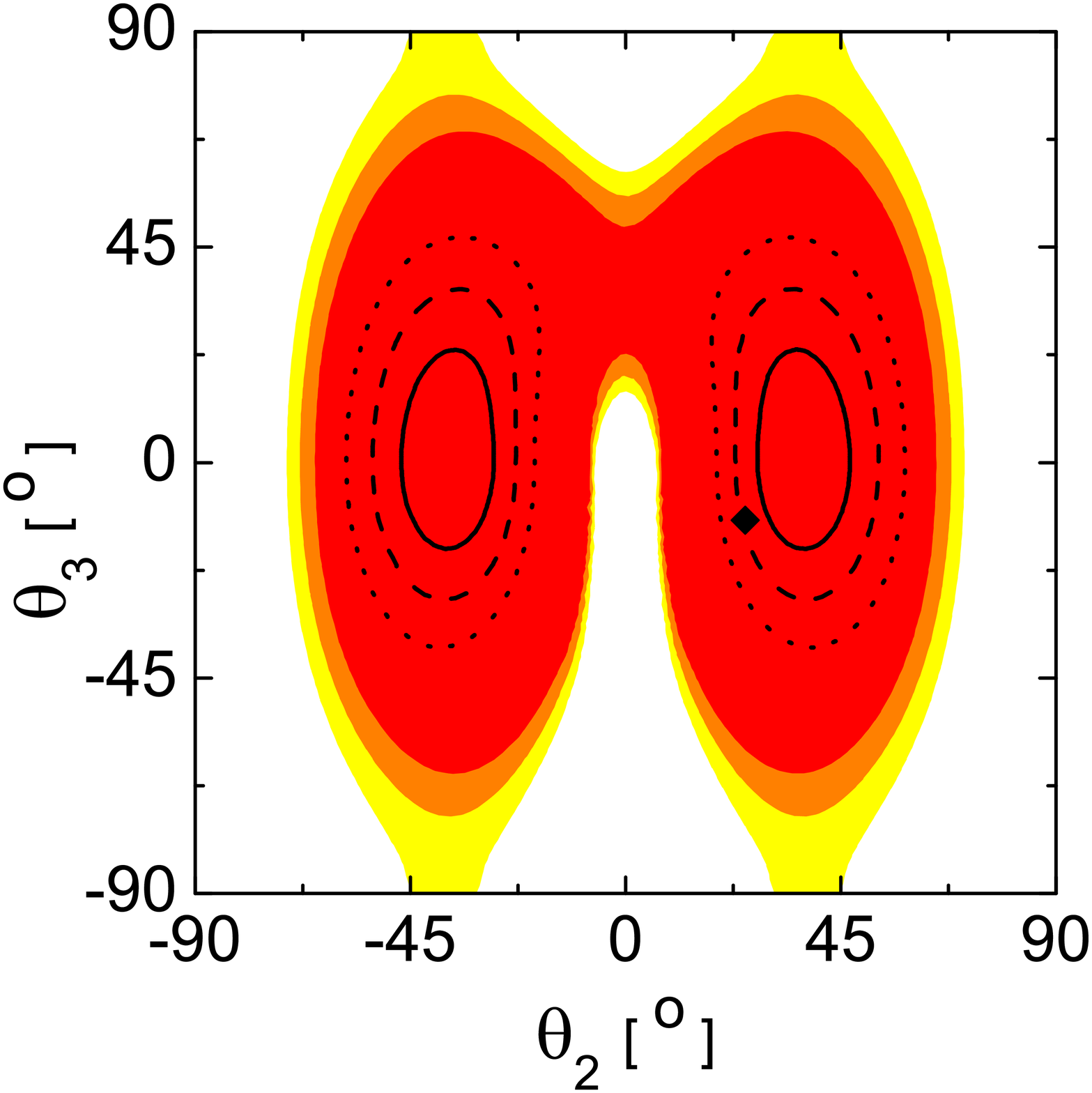}
\caption{\label{fig:fig4} Parameter spaces of $\vartheta_i$ relating
the bi-maximal mixing pattern at the scale $\mu = M_3$ to the
experimentally measured leptonic mixing angles at 1$\sigma$ (red
area, solid curve), 2$\sigma$ (orange area, dashed curve), and
3$\sigma$ (yellow area, dotted curve) C.L. The colored areas
correspond to the choice $m_1(M_3)=0.05~{\rm eV}$, while the curves
$m_1(M_3)=0.03~{\rm eV}$. The black diamonds represent the input
values of $\vartheta_i$ in Figs.~\ref{fig:fig2} and \ref{fig:fig3}.}
\vspace{0cm}
\end{center}
\end{figure}

As a result of the freedom in $O$, various RG running behaviors of
the leptonic mixing angles become viable for different choices of
$\vartheta_i$. Once a certain mixing pattern is fixed at a high
energy scale, i.e. $\mu=M_3$, one can tune the rotation angles
$\vartheta_i$ so as to achieve the experimentally measured data at
low energy scales. Concretely, we show an example in
Fig.~\ref{fig:fig4}, in which the bi-maximal mixing pattern of $U$
is fixed at the scale $\mu = M_3$, while the parameter spaces of
$\vartheta_i$ at 1$\sigma$, 2$\sigma$, and 3$\sigma$ C.L. are shown
in the plot. The colored areas correspond to the choice
$m_1(M_3)=0.05~{\rm eV}$, while the curves $m_1(M_3)=0.03~{\rm eV}$.
As we have estimated before, in the case of $m_1=0.05~{\rm eV}$,
significant corrections could be received for $\theta_{12}$, and
therefore, the bi-maximal mixing pattern can be easily fetched at
higher energy scales. For the case of a smaller $m_1$, the parameter
spaces of $\vartheta_i$ shrink a lot, and a larger value of
$\vartheta_1$, e.g., $\vartheta_1 \simeq 90^\circ$, is more favorable.
In particular, $\vartheta_1 \simeq 0$ is ruled out, which can
be seen from Eq.~\eqref{eq:th12} where the threshold correction to
$\theta_{12}$ is suppressed for small $\vartheta_1$.

We remark that, although we only consider the normal hierarchy of
light neutrino masses in our numerical analysis, the main results
are unchanged if the inverted mass hierarchy is assumed, i.e.,
$m_3<m_1<m_2$. The reason is simply that the RG running of
$\theta_{12}$ is mainly affected by the inverse of $\Delta
m^2_{21}$, which does not flip its sign for different neutrino mass
hierarchies. In the limit of a highly degenerate spectrum of light
neutrino masses, i.e., $m_1\simeq m_2\simeq m_3$, sizable
enhancements of RG running can also be acquired for both
$\theta_{13}$ and $\theta_{23}$, of which the light neutrino mass
hierarchies may play a key role in the RG running. Furthermore, in
our numerical calculations, we have fixed the mass spectrum of the
right-handed neutrinos, in which the interval between two such
neutrinos is around a few hundred GeV. In general, a stronger
hierarchy among the masses of the right-handed neutrinos can lead to
more sizable RG corrections to the neutrino parameters, whereas a
weakly hierarchical mass spectrum of the right-handed neutrinos
would suppress the seesaw threshold effects.

\section{Summary and conclusions}
\label{sec:summary}

In this work, we have studied the RG running of neutrino parameters
in the low-scale type-I seesaw model. We have shown that, in the RG
evolution of the leptonic mixing angles between the seesaw
thresholds, significant radiative corrections could be obtained even
for a short distance of RG running. In particular, we have presented
analytical formulas for the RG corrections to the neutrino
parameters in crossing the seesaw thresholds, which clearly indicate
that sizable threshold corrections to the leptonic mixing matrix may
occur due to the mismatch between different contributions to the
mass matrix of the light neutrinos. In addition, we have
demonstrated that, if the light neutrino mass spectrum is nearly
degenerate, there exist large enhancement factors boosting the RG
running of the leptonic mixing angles remarkably. A numerical
example has also been given to show that, in the presence of
low-scale right-handed neutrinos, the bi-maximal mixing pattern at
TeV scale is fully compatible with the current measured leptonic
mixing angles. This peculiar feature turns out to be very useful for
the construction of a low energy scale flavor symmetric theory, and
in particular, it provides us with the possibility of testing such a
flavor symmetry at running and forthcoming experiments (such as the
LHC). For the sake of keeping the seesaw scale within the reach of
these experiments, the neutrino Yukawa coupling tends to be chosen
very small in order to stabilize the mass scale of the light
neutrinos. One might think that this prevents the low-scale type-I
seesaw model from being testable, since the predominant interaction
between the right-handed neutrinos and the SM fields is through the
Yukawa coupling. However, this is not necessarily true in the
presence of additional interactions. For example, if the
right-handed neutrinos and other SM fields are charged under an
additional gauge group, broken not far above the TeV scale, such a
new gauge interaction could give rise to observable effects. Note
that, although we mainly concentrate on the type-I seesaw model
throughout this work, our main conclusions can be easily extended to
various low-scale seesaw models. For example, in the type-I+II
seesaw framework with three right-handed neutrinos and a heavy Higgs
triplet field being introduced, both the right-handed neutrinos and
the Higgs triplet may contribute to the masses of the light
neutrinos, and the mismatch between different contributions would
result in similar threshold corrections to the leptonic mixing
angles.

\begin{acknowledgments}

One of the authors (H.Z.) would like to thank Zhi-zhong Xing for
valuable discussions especially during the initial stage of this
work.

This work was supported by the Swedish Research Council
(Vetenskapsr{\aa}det), contract No. 621-2008-4210 (T.O.) and the
Royal Institute of Technology (KTH), project No. SII-56510 (H.Z.).

\end{acknowledgments}

\newpage

\appendix
\section{Derivation of the analytical formulas for the corrected leptonic mixing matrix and the light neutrino masses}
\label{sec:app-A}

In this appendix, we derive the first-order threshold corrections to
the leptonic mixing matrix and the light neutrino masses. In
general, the light neutrino mass matrix is complex and symmetric,
i.e., $m_\nu = m^T_\nu$. In order to derive the series
expansions, we write the perturbed mass matrix as
\begin{eqnarray}
\tilde{m}_{\nu} =m_{\nu} + m'_{\nu},
\end{eqnarray}
where $m'_{\nu}$ corresponds to a correction to $m_\nu$. The
corrected leptonic mixing matrix is given by $\tilde{U} = U + U'$
with $U'$ being the correction of $U$. In analogy, one can define
the real and diagonal matrix $\tilde{D} = D + D'$, i.e., $\tilde{D}=
{\rm diag} (\tilde{m_1}, \tilde{m_2}, \tilde{m_3})$, and similarly
for the matrices $D$ and $D'$.

By definition, we have
\begin{eqnarray}
\tilde{U}\tilde{D}\tilde{U}^T = \tilde{m}_{\nu}, \quad U
DU^T = m_{\nu},
\end{eqnarray}
and to lowest order in the corrections $U'$ and $D'$, we find that
\begin{eqnarray}
m'_{\nu} = \tilde{m}_{\nu} - m_{\nu} = \tilde{U}\tilde{D}\tilde{U}^T
-  U DU^T \simeq UDU'^T + U'DU^T + UD'U^T.
\end{eqnarray}
Multiplying with $U^{\dagger}$ from the left and $U^*$ from the
right to the above equation, we obtain
\begin{eqnarray}
U^{\dagger}m'_{\nu}U^* = DU'^TU^* + U^{\dagger}U'D + D'.\label{eq:correction}
\end{eqnarray}
Now, we define an anti-Hermitian matrix $T$ and a symmetric matrix
$\Lambda$ as
\begin{eqnarray}
T & = & U^\dagger U', \label{eq:def-T} \\
\Lambda & = &  U^{\dagger}m'_{\nu}U^*, \label{eq:def-Lambda}
\end{eqnarray}
Here one can easily prove that $T^{\dagger} = - T$, and the diagonal
elements of $T$ are purely imaginary, i.e., $T_{ii}=-T^*_{ii}$.
Using Eqs.~(\ref{eq:correction})-(\ref{eq:def-Lambda}), $\Lambda$
can be computed as
\begin{eqnarray}
\Lambda =  -D T^* + TD + D'. \label{eq:correq}
\end{eqnarray}
In order to obtain the corrections to the mass eigenvalues, we
investigate the diagonal part of this matrix equation, which
indicates that (no summation over repeated indices)
\begin{eqnarray}
\Lambda_{ii} = -m_{i}T^*_{ii}+T_{ii}m_i +m'_i,
\end{eqnarray}
and the real part gives the corrections to the neutrino masses,
i.e.,
\begin{eqnarray}
m_i' = \Re (\Lambda_{ii}).\label{eq:mi_pert}
\end{eqnarray}
In addition, the imaginary part is
\begin{eqnarray}
\Im (T_{ii}) = \frac{\Im (\Lambda_{ii})}{2m_i}. \label{eq:Tii}
\end{eqnarray}
Next, to obtain the corrections to the leptonic mixing matrix, we
study the non-diagonal part of Eq.~(\ref{eq:correq}), which is
\begin{eqnarray}
\Lambda_{ij} = -m_i T^*_{ij} + T_{ij}m_j, \quad i \neq j.
\label{eq:lambda-ij}
\end{eqnarray}
Now, the real and imaginary parts of Eq.~\eqref{eq:lambda-ij} yield
\begin{eqnarray} \Re (T_{ij}) &=& \frac {\Re (\Lambda_{ij})}{m_j-m_i}, \label{eq:TijRe} \\
\Im (T_{ij}) &=& \frac {\Im (\Lambda_{ij})}{m_i+m_j},\label{eq:TijIm}
\end{eqnarray}
respectively. Therefore, all the elements of $T$ can be obtained by
using Eqs.~(\ref{eq:Tii}), (\ref{eq:TijRe}) and (\ref{eq:TijIm}),
and $U'$ can be calculated according to Eq.~\eqref{eq:def-T}.

Note that, if $m'_{\nu} = \xi m_{\nu}$, where $\xi$ is a constant,
then $\tilde{D} = (1+\xi)D$ and $T=0$, and thus, $\tilde{U} = U$.
This means that, if $m'_{\nu}$ is proportional to $m_{\nu}$, $U$
will not be affected. As a special case, consider CP conservation,
when the original mixing matrix $U$ as well as the perturbation
$m_{\nu}'$ are real. Then, $\Lambda^* = \Lambda$, and therefore, one
has
\begin{eqnarray}
m'_i &=& \Lambda_{ii}, \\
T_{ii} &=& 0, \label{eq:corrections_CPC}\\
T_{ij} &=&  \frac {\Lambda_{ij}}{m_j-m_i}, \quad i \neq j.
\end{eqnarray}

As one application, we consider the threshold corrections in the
main context. Below the scale $\mu = M_n$, one should take [cf.~Eq.~(\ref{eq:mismatch})]
\begin{eqnarray}
m'_{\nu} = (b-1)m_{\nu} + \varepsilon v^2 \kappa^{(n)},
\end{eqnarray}
as the perturbation to $m_\nu$. Using the matching condition at
$\mu=M_n$, $\kappa^{(n)}$ can be expressed in terms of $Y_{\nu}$ as
\begin{eqnarray}
\kappa^{(n)} =\frac{\mathcal{Y}_{\nu}^{(n)}  \mathcal{Y}_{\nu}^{(n)T}}{M_n},
\end{eqnarray}
where $\mathcal{Y}_{\nu}^{(n)}$ is the $n$-th column of $Y_{\nu}$.
Here we will make use of the Casas--Ibarra parametrization of
$Y_\nu$ [cf., Eq.~\eqref{eq:CI}], which allows us to express the
corrections to $U$ in terms of the rotation angles $\vartheta_i$ and
the low-energy parameters. For a general $U$ and $Y_{\nu}$, one can
calculate the matrix $\Lambda = \Lambda^{(n)}$, giving the
contribution of $\kappa^{(n)}$ for $\mu$ below the threshold
$M_n$,\footnote{If all the right-handed neutrinos are located at the
same energy scale, one would effectively obtain a summation over the
index $n$, giving the result $\Lambda_{ij} = (b-1+\varepsilon) \delta_{ij} m_i = a
\delta_{ij} m_i $, since $OO^T=1$.}
\begin{eqnarray}\label{eq:Lambda_calc}
\Lambda^{(n)}_{ij}=\left(U^{\dagger}m'_{\nu} U^* \right)_{ij} =
(b-1) \delta_{ij} m_i + \varepsilon O_{i n}O_{j n} \sqrt{m_i m_j}.
\end{eqnarray}
Inserting the above equation into Eq.~(\ref{eq:mi_pert}), one
directly arrives at the neutrino masses at the scale $\mu=M_{n-1}$,
i.e.,
\begin{eqnarray}\label{eq:general_mi}
\tilde{m}^{}_i &=&  \left[ b + \varepsilon \Re(O_{in}^2) \right] m_i. \label{eq:tildem1}
\end{eqnarray}
Also, the components of the matrix $T$ can be calculated as
\begin{eqnarray}\label{eq:general_T}
T^{(n)}_{ii} &=& \frac{{\rm i} \varepsilon}{2} \Im(O_{in}^2),\\
T^{(n)}_{ij} &=&  \varepsilon \sqrt{m_i m_j} \left[
\frac{\Re(O_{in}O_{jn})}{m_j-m_i} +{\rm i}
\frac{\Im(O_{in}O_{jn})}{m_i+m_j}\right], \quad i \neq j.
\end{eqnarray}
Note that $O_{ij}$ may, in principle, have very large absolute
values if they are complex. If $O$ is real, or purely
imaginary,\footnote{That is with ``low-energy'' but without
``high-energy'' CP-violation.} the corrected mixing matrix at the lower
scale $\mu=M_{n-1}$ reads
\begin{eqnarray}
\label{eq:U_corr}\tilde{U}_{\alpha j} = U_{\alpha j} +
\left[UT^{(n)}\right]_{\alpha j} =  U_{\alpha j} +\varepsilon
\sum_{k\neq j} \frac{\sqrt{m_j m_k} }{m_j-m_k} O_{kn}O_{jn}
U_{\alpha k}.
\end{eqnarray}

\section{Corrections to the leptonic mixing angles and the light neutrino masses}
\label{sec:app-B}

In the following, we use the analytical results obtained in
Appendix~\ref{sec:app-A} to derive explicit expressions for the
threshold corrections to the leptonic mixing angles and the light
neutrino masses. Here, for simplicity, we assume CP is preserved,
i.e., all the CP-violating phases are taken to be zero.
Since the perturbation is linear in $\kappa$, one should, for $\mu$
between the $n$-th and $(n-1)$-th thresholds, simply add the
contributions from all the thresholds above $\mu$.

\subsection{Threshold corrections between $M_3$ and $M_2$}

\subsubsection{Correction to $\theta_{13}$}

Using Eq.~(\ref{eq:U_corr}) with $n=3$, the corrected quantity to
$s_{13}$ is given by
\begin{eqnarray}
s'_{13}=\tilde{s}_{13} -s_{13} &=& \tilde{U}_{e3} - U_{e3} =
\sum_{k\neq 3} U_{e k}T^{(3)}_{k3} =\varepsilon \sum_{k\neq 3}
\frac{\sqrt{m_3 m_k} }{m_3-m_k} O_{k3}O_{33} U_{e k} \nonumber \\
&=& \varepsilon\frac{\sqrt{m_1 m_3} O_{13} O_{33}}{m_3-m_1}
c_{13}c_{12} +\varepsilon\frac{\sqrt{m_2 m_3} O_{23}
O_{33}}{m_3-m_1} c_{13}s_{12}\nonumber\\ &=& \varepsilon
c_{{1}}c_{{2}}c_{{13}} \sqrt{m_{{3}}} \left( {\frac
{s_{{2}}c_{{12}}\sqrt { m_{{1}}}}{m_{{3}}-m_{{1}}}} + {\frac
{s_{{1}}c_{{2}}s_{{12}}\sqrt{m_{{2}}}}{m_{{3}}-m_{{2}}}}\right).
\end{eqnarray}
Taylor expanding this equation around $\tilde s_{13} = s_{13}$, one
obtains
\begin{eqnarray}
\tilde{\theta}_{13} &=& \arcsin(s_{13} +s'_{13}) =
\theta_{13}+\frac{s'_{13}}{c_{13}} +
\mathcal{O}\left[(s'_{13})^2\right] \nonumber\\
&\simeq&
\theta_{13}+\varepsilon c_{{1}}c_{{2}} \sqrt {m_{{3}}} \left( {\frac
{s_{{1}}c_{{2}}s_{{12}}\sqrt {m_{{2}}}}{m_{{3}}-m_{{2}}}}+{\frac
{s_{{2}}c_{{12}}\sqrt { m_{{1}}}}{m_{{3}}-m_{{1}}}} \right).
\end{eqnarray}
In the nearly degenerate case, i.e., $m_1\simeq m_2 \simeq m_3$, the
above formula approximates to
\begin{eqnarray}
\theta'_{13} = \tilde{\theta}_{13} - \theta_{13} \simeq \varepsilon
{\frac { c_{{1}}c_{{2}}\left( s_{{2}}c_{{12}} + s_{{1}}c_{{2}}
s_{{12}} \right) \sqrt{m_{{1}}m_{{3}}} }{m_{{3}}-m_{{1} }}}.
\end{eqnarray}
One can observe that the correction to $\theta_{13}$ is actually
independent of $s_{13}$. Furthermore, note that this expression is
proportional to $\sqrt{m_1 m_3}/(m_3-m_1)$. Thus, the correction
cannot be very large, since there is no $1/(m_2-m_1)$ enhancement.

\subsubsection{Correction to $\theta_{12}$}

Similarly, using Eq.~(\ref{eq:U_corr}), the threshold correction
to $U_{e2}$ can be obtained as
\begin{eqnarray}
U'_{e2}&=& \tilde{U}_{e2} - U_{e2} = \sum_{k\neq 2} U_{e
k}T^{(3)}_{k2} = \varepsilon \sum_{k\neq 2} \frac{\sqrt{m_2 m_k}
}{m_2-m_k} O_{k3}O_{23} U_{e k} \nonumber\\
&=& \varepsilon s_1c_2
\sqrt{m_2} \left( \frac{s_2c_{12}c_{13}\sqrt {m_ 1}}{m_2-m_1} -
{\frac {c _1c_2 s_{13}\sqrt {m_3}}{m_3-m_2}} \right).
\end{eqnarray}
To obtain the correction to $s_{12}$, we note that, in the standard
parametrization,
\begin{eqnarray}
\tilde{s}_{12} = \frac{\tilde{U}_{e2}}{\tilde{c}_{13}}.
\end{eqnarray}
Expanding in the small correction $s'_{13}$, we have
\begin{eqnarray}
\frac{1}{\tilde{c}_{13}} \simeq
\frac{1}{c_{13}}\left(1+\frac{s_{13}s'_{13}}{c^2_{13}} \right),
\end{eqnarray}
and therefore, we find that
\begin{eqnarray}
\tilde{s}_{12} =
\left(U_{e2}+U'_{e2}\right)\frac{1}{c_{13}}\left(1+\frac{s_{13}s'_{13}}{c^2_{13}}
\right) \simeq s_{12} + s_{12} \frac{s_{13}s'_{13}}{c^2_{13}} +
\frac{U'_{e2}}{c_{13}},
\end{eqnarray}
where the corrections $s'_{13}$ and $U'_{e2}$ have been calculated
above. For $s_{13} \simeq 0$, one obtains
\begin{eqnarray}
s'_{12} = \tilde{s}_{12} -  s_{12} \simeq  \varepsilon s_1 s_2 c_2
c_{12} \frac{\sqrt{m_1 m_2}}{m_2-m_1}.
\end{eqnarray}
In addition, Taylor expanding around $\tilde{s}_{12} = s_{12}$ gives
\begin{eqnarray}
\theta'_{12} = \tilde{\theta}_{12} - \theta_{12} \simeq \varepsilon
s_1 s_2 c_2 \frac{\sqrt{m_1 m_2}}{m_2-m_1},
\label{eq:theta'12}
\end{eqnarray}
where the right-hand side is independent of $\theta_{12}$.

\subsubsection{Correction to $\theta_{23}$}

Finally, using Eq.~(\ref{eq:U_corr}), the threshold correction to $U_{\mu
3}$ is given by
\begin{eqnarray}
U'_{\mu 3}  &=& \tilde{U}_{\mu 3} - U_{\mu 3} = \sum_{k\neq 3}
U_{\mu k}T^{(3)}_{k3} = \varepsilon \sum_{k\neq 3} \frac{\sqrt{m_3
m_k} }{m_3-m_k} O_{k3}O_{33} U_{\mu k} \nonumber \\
&=&  \varepsilon c_{{1}}c_{{2}} \sqrt {m_{{3}}} \left(
\frac{s_{{2}}U_{\mu 1} \sqrt{m_{{1}}}}{m_{{3}}-m_{{1}}} +
\frac{s_{{1}}c_{{2}} U_{\mu
2}\sqrt{m_{{2}}}}{m_{{3}}-m_{{2}}}\right).
\end{eqnarray}
As for the corrections to $s_{23}$, we note that, in the standard
parametrization,
\begin{eqnarray}
\tilde{s}_{23} = \frac{\tilde{U}_{\mu 3}}{\tilde{c}_{13}},
\end{eqnarray}
and thus, one obtains
\begin{eqnarray}
\tilde{s}_{23} = \left(U_{\mu 3}+U'_{\mu
3}\right)\frac{1}{c_{13}}\left(1+\frac{s_{13}s'_{13}}{c^2_{13}}
\right) \simeq s_{23} + s_{23} \frac{s_{13}s'_{13}}{c^2_{13}} +
\frac{U'_{\mu 3}}{c_{13}},
\end{eqnarray}
where $s'_{13}$ and $U'_{\mu 3}$ have been calculated above. For
$s_{13} \simeq 0$, one obtains
\begin{eqnarray}
s'_{23}=\tilde{s}_{23} - s_{23} \simeq \varepsilon
c_{{1}}c_{{2}}c_{{23}} \sqrt {m_{{3}}} \left( - s_{{2}} s_{{12}}
\frac{\sqrt{m_{{1}}}}{m_{{3}}-m_{{1}}}  + s_{{1}}c_{{2}} c_{{12}}
\frac{\sqrt{m_{{2}}}}{m_{{3}}-m_{{2}}} \right),
\end{eqnarray}
giving that
\begin{eqnarray}
\theta'_{23} = \tilde{\theta}_{23} - \theta_{23} \simeq \varepsilon
c_{{1}}c_{{2}} \sqrt {m_{{3}}} \left(- s_{{2}} s_{{12}}
\frac{\sqrt{m_{{1}}}}{m_{{3}}-m_{{1}}} + s_{{1}}c_{{2}} c_{{12}}
\frac{\sqrt{m_{{2}}}}{m_{{3}}-m_{{2}}}  \right).
\end{eqnarray}
One finds that $\theta'_{23}$ is independent of $\theta_{23}$. In
the nearly degenerate limit, the above formula approximates to
\begin{eqnarray}
\theta'_{23} = \varepsilon c_{{1}}c_{{2}}\left(s_1 c_2 c_{12} -s_2 s_{12} \right) \frac{\sqrt {m_1 m_3}}{m_3 -
m_1}.
\end{eqnarray}
Similar to the corrections to $\theta_{13}$, there is no strong
enhancement factor.

\subsubsection{Corrections to the light neutrino masses}

According to Eq.~(\ref{eq:general_mi}), the explicit expressions for
the threshold corrections to the light neutrino masses are
\begin{eqnarray}
m_1' &=& \tilde{m}_1 - m_1  =  \left[ (b -1)+ \varepsilon s_2^2 \right]m_1, \\
m_2' &=& \tilde{m}_2 - m_2  =  \left[ (b -1)+ \varepsilon s_1^2 c_2^2 \right]m_2, \\
m_3' &=& \tilde{m}_3 - m_3  =  \left[ (b -1)+ \varepsilon c_1^2
c_2^2 \right]m_3.
\end{eqnarray}

\subsection{Threshold corrections between $M_2$ and $M_1$}

One may repeat the above analysis with $n=2$, and calculate the
threshold corrections to the neutrino parameters between $M_2$ and
$M_1$. Explicitly, for $s_{13}$, one obtains
\begin{eqnarray}
s'_{1 3} &=& \varepsilon O_{32} c_{13} \sqrt{m_3} \left( {\frac {c_2
s_3 c_{12}\sqrt { m_1}}{m_3-m_1}} + {\frac {O_{22} s_{12}
\sqrt{m_2}}{m_3-m_2}}\right),
\end{eqnarray}
and henceforth,
\begin{eqnarray}
\theta'_{13} =  \varepsilon O_{32} \sqrt{m_3} \left( {\frac {c_2 s_3
c_{12}\sqrt { m_1}}{m_3-m_1}} + {\frac {O_{22} s_{12}
\sqrt{m_2}}{m_3-m_2}}\right),
\end{eqnarray}
where $O_{22} = c_1 c_3 -s_1 s_2 s_3$ and $O_{32}=-s_1 c_3 - c_1 s_2
s_3$. In the nearly degenerate limit, the above expression
approximates to
\begin{eqnarray}
\theta'_{13} \simeq \varepsilon O_{32} \frac{\sqrt{m_1
m_3}}{m_3-m_1} \left( c_2 s_3 c_{12} +  O_{22} s_{12}\right).
\end{eqnarray}
Then, for $\theta_{12}$, one obtains
\begin{eqnarray}
U'_{e 2} = \varepsilon O_{22} \sqrt{m_2} \left(
\frac{c_2s_3c_{12}c_{13}\sqrt {m_ 1}}{m_2-m_1} - {\frac {O_{32}
s_{13}\sqrt {m_3}}{m_3-m_2}} \right).
\end{eqnarray}
Ignoring the small parameter $s_{13}$, we arrive at
\begin{eqnarray}
s'_{12} \simeq \varepsilon c_2s_3\left(c_1c_3-s_1s_2s_3  \right)c_{12} \frac{\sqrt{m_1
m_2}}{m_2-m_1},
\end{eqnarray}
which corresponds to
\begin{eqnarray}
\theta'_{12} \simeq \varepsilon c_2s_3\left(c_1c_3-s_1s_2s_3  \right) \frac{\sqrt{m_1
m_2}}{m_2-m_1}.
\label{eq:theta'12_2}
\end{eqnarray}
Next, for $\theta_{23}$, the correction to $U_{\mu 3}$ reads
\begin{eqnarray}
U'_{\mu 3} = \varepsilon O_{32} \sqrt{m_3} \left( \frac{c_2s_3U_{\mu
1}\sqrt{m_1}}{m_3-m_1} + \frac{O_{22} U_{\mu 2}\sqrt{m_2}}{m_3-m_2}
\right),
\end{eqnarray}
which, for $s_{13} \simeq 0$ and in the nearly degenerate case, gives
\begin{eqnarray}
s'_{23} \simeq \varepsilon O_{32}\left( -c_2s_3s_{12} +
O_{22}c_{12}\right) c_{23}
\frac{\sqrt{m_1m_3}}{m_3-m_1},
\end{eqnarray}
and
\begin{eqnarray}
\theta'_{23} \simeq \varepsilon O_{32}\left(
-c_2s_3s_{12} + O_{22}c_{12}\right) \frac{\sqrt{m_1m_3}}{m_3-m_1}.
\end{eqnarray}
Finally, the threshold corrections to the light neutrino masses are
given by
\begin{eqnarray}
m_1' &=& \left[ (b -1) + \varepsilon c_2^2 s_3^2 \right]m_1, \\
m_2' &=&  \left[ (b -1)+ \varepsilon \left( c_1 c_3-s_1 s_2 s_3 \right)^2\right] m_2, \\
m_3' &=&  \left[ (b -1) + \varepsilon \left(s_1 c_3 +  c_1 s_2
s_3\right)^2\right] m_3.
\end{eqnarray}


\end{document}